\global\def\draftcontrol{0}

   \def\versionno{Quant geometry}

\catcode`\@=11

\expandafter\ifx\csname draftcontrol\endcsname\relax\global\def\draftcontrol{0} 
\fi 

{\count255=\time\divide\count255 by 60 
\xdef\hourmin{\number\count255} 
\multiply\count255 by-60\advance\count255 by\time 
\xdef\hourmin{\hourmin:\ifnum\count255<10 0\fi\the\count255}} 
\def\draftdate{\number\month/\number\day/\number\year\ \ \ \hourmin } 

\newcommand\makepapertitle{\par

  \begingroup 
    \renewcommand\thefootnote{\@fnsymbol\c@footnote}%
    \def\@makefnmark{\rlap{\@textsuperscript{\normalfont\@thefnmark}}}%
    \long\def\@makefntext##1{\parindent 1em\noindent 
            \hb@xt@1.8em{%
                \hss\@textsuperscript{\normalfont\@thefnmark}}##1}%
     \newpage 
     \global\@topnum\z@   
     \@makepapertitle 
     \thispagestyle{empty}\@thanks 
  \endgroup 
  \setcounter{footnote}{0}%
  \global\let\thanks\relax 
  \global\let\makepapertitle\relax 
  \global\let\@makepapertitle\relax 
  \global\let\@thanks\@empty 
  \global\let\@author\@empty 
  \global\let\@date\@empty 
  \global\let\@title\@empty 
  \global\let\title\relax 
  \global\let\author\relax 
  \global\let\date\relax 
  \global\let\and\relax 
  \def\version{\let\version\@version\@gobble} 
} 
\def\@makepapertitle{%
  \newpage 
   \ifnum\draftcontrol=1 {} 
   \version\versionno 
   \vskip 5.5em%
   \else 
   \hfill\hbox to 3cm {\parbox{4.5cm}{\@pubnum}\hss}%
   \vskip 6.5em%
   \fi 
   \begin{center}%
   \let \footnote \thanks 
      {\hskip -0\textwidth \hbox to 1\textwidth%
        {\centerline{\Large\bf{\noindent\@title}}}}%
     \vskip 2em%
     {\normalsize
       \lineskip .5em%
       \begin{tabular}[t]{c}%
         \@author 
       \end{tabular}\par}%
     \vskip 1.5em%
     {\@bstract}%
     \end{center}%
     \vfill
     \@date%
     \vskip 1.5em%
   \par 
} 

\gdef\@pubnum{} 
\def\pubnum#1{%
  \gdef\@pubnum{#1}} 

\gdef\@bstract{} 
\def\Abstract#1{%
  \gdef\@bstract{%
   \parbox{\textwidth-0pc}{%
   \centerline{\bf Abstract}\penalty1000 
   \noindent
   \renewcommand\baselinestretch{1.0} 
   {#1}}} 
} 

\gdef\@email{}
\def\email#1{%
   \gdef\@email{%
   Email: {\tt #1}}
}

\def\ps@paper{\let\@mkboth\@gobbletwo%
     \ifnum\draftcontrol=1 
        \def\@oddfoot{\hbox to \textwidth{\tiny \versionno \hfil\tiny\draftdate}%
        \hskip -\textwidth \hbox to \textwidth{\hfil\rm\thepage\hfil}}%
     \else\def\@oddfoot{\hbox to \textwidth{\hfil\rm\thepage\hfil}} 
     \fi 
     \let\@evenfoot\@oddfoot 
} 

\def\body{\clearpage 
          \pagestyle{paper} 
        } 
\newenvironment{acknowledgments}{%
\vskip 3.25ex 
\addcontentsline{toc}{section}{Acknowledgments}
\noindent {\bf Acknowledgments} 
} 

\def\@version#1{\ifnum\draftcontrol=1 
\typeout{}\typeout{#1}\typeout{} 
\vskip3mm\centerline{\hbox{\fbox{\normalsize{\tt DRAFT -- #1 -- } 
                   {\draftdate}}}}\vskip3mm 
\fi} 
\let\version\@version 
\long\def\eqlabel#1{\ifnum\draftcontrol=1 
                    \tag@false  
                    \tag*{(\theequation) \hbox to -0.2cm{\hspace{0cm}\small{#1}\hss}} 
                    \refstepcounter{equation}  
                    \edef\@currentlabel{\theequation} 
                    \ltx@label{#1}          
                    \else 
                    \label{#1} 
                    \fi 
                    } 
\let\st@bibitem\@bibitem 
\let\st@lbibitem\@lbibitem 
\ifnum\draftcontrol=1 
  \def\@bibitem#1{%
    \st@bibitem{#1}\a@@label{#1}\ignorespaces} 
  \def\@lbibitem[#1]#2{%
    \st@lbibitem[#1]{#2}\a@@label{#2}\ignorespaces} 
  \def\a@@label#1{%
    \gdef\a@lab{\smash{\normalfont\small#1}} 
    \ifvmode 
      \if@inlabel 
        \global\setbox\@labels\hbox{%
          \llap{\a@lab\let\a@lab\relax 
                \kern\@totalleftmargin\kern\marginparsep}%
          \box\@labels}%
      \fi 
    \fi} 
\fi 

\documentclass[12pt,letterpaper]{article} 

\usepackage{amsmath,bm,amsfonts,amssymb,array,calc,amsthm,rotating,amscd}
\usepackage{epsfig,psfrag} 
\usepackage{rotating}
\usepackage{amscd}
\usepackage{graphicx}
\usepackage{color}
\usepackage[colorlinks=false]{hyperref}

\renewcommand\baselinestretch{1.25} 
\renewcommand\section{\@startsection {section}{1}{\z@}%
                                   {-3.5ex \@plus -1ex \@minus -.2ex}%
                                   {2.3ex \@plus.2ex}%
                                   {\normalfont\large\bfseries}} 
\renewcommand\subsection{\@startsection{subsection}{2}{\z@}%
                                   {-3.25ex\@plus -1ex \@minus -.2ex}%
                                   {1.5ex \@plus .2ex}%
                                   {\normalfont\normalsize\bfseries}} 
\renewcommand\subsubsection{\@startsection{subsubsection}{3}{\z@}%
                                   {-3.25ex\@plus -1ex \@minus -.2ex}%
                                   {1.5ex \@plus .2ex}%
                                   {\normalfont\normalsize\it}} 
\renewcommand\paragraph{\@startsection{paragraph}{4}{\z@}%
                                   {-1.75ex\@plus -1ex \@minus -.2ex}%
                                   {1ex \@plus .2ex}%
                                   {\normalfont\normalsize\bf}} 
\renewcommand\subparagraph{\@startsection{subparagraph}{5}{\z@}%
                                   {-1.25ex\@plus -0ex \@minus -.2ex}%
                                   {-2ex \@plus .2ex}%
                                   {\normalfont\normalsize\it}}

\numberwithin{equation}{section}

\long\def\@makecaption#1#2{%
  \vskip\abovecaptionskip
  \sbox\@tempboxa{{\bf #1:} #2}%
  \ifdim \wd\@tempboxa >\hsize
    {\small\bf #1:} {\small #2}\par
  \else
    \global \@minipagefalse
    \hb@xt@\hsize{\hfil\box\@tempboxa\hfil}%
  \fi
  \vskip\belowcaptionskip}

\renewcommand*\l@section[2]{%
  \ifnum \c@tocdepth >\z@
    \addpenalty\@secpenalty
    \addvspace{.5em \@plus\p@}%
    \setlength\@tempdima{1.5em}%
    \begingroup
      \parindent \z@ \rightskip \@pnumwidth
      \parfillskip -\@pnumwidth
      \leavevmode \bfseries
      \advance\leftskip\@tempdima
      \hskip -\leftskip
      #1\nobreak\hfil \nobreak\hb@xt@\@pnumwidth{\hss #2}\par
    \endgroup
  \fi}
\renewcommand*\l@subsection{\addvspace{.0em \@plus\p@}\@dottedtocline{2}{1.5em}{2.3em}}
\renewcommand*\l@subsubsection{\addvspace{-.2em \@plus\p@}\@dottedtocline{3}{3.8em}{3.2em}}

\def\quantph#1{\href{http://xxx.arxiv.org/abs/quant-ph/#1}{{arXiv:quant-ph/#1}}}
\def\hepth#1{\href{http://xxx.arxiv.org/abs/hep-th/#1}{{arXiv:hep-th/#1}}}

\def\hepph#1{\href{http://xxx.arxiv.org/abs/hep-ph/#1}{{arXiv:hep-ph/#1}}}

\def\arxiv#1#2{\href{http://xxx.arxiv.org/abs/#1}{{arXiv:#1 [#2]}}}

\definecolor{refcol}{rgb}{0.2,0.2,0.8}
\definecolor{eqcol}{rgb}{.6,0,0}
\definecolor{purple}{cmyk}{0,1,0,0}

\gdef\@citecolor{refcol}
\gdef\@linkcolor{eqcol}
\def\colorlinkspurple{\gdef\@urlcolor{purple}}
\def\colorlinksblue{\gdef\@urlcolor{blue}}
\def\colorlinksred{\gdef\@urlcolor{red}}

\def\ie{{\it i.e.}}

\def\cf{{\it cf.}}

\def\revise#1       {\raisebox{-0em}{\rule{3pt}{1em}}%
                     \marginpar{\raisebox{.5em}{\vrule width3pt\ 
                     \vrule width0pt height 0pt depth0.5em 
                     \hbox to 0cm{\hspace{0cm}{%
                     \parbox[t]{4em}{\raggedright\footnotesize{#1}}}\hss}}}}

\def\calb         {{\cal B}}

\def\cale         {{\cal E}}

\def\ii           {{\it i}}

\def\sqr#1#2{{\vcenter{\vbox{\hrule height.#2pt   
 \hbox{\vrule width.#2pt height#1pt \kern#1pt 
 \vrule width.#2pt}\hrule height.#2pt}}}}

\renewcommand{\S}{\mathbb S}

\newcommand{\Fcal}{\mathcal F}

\newcommand{\Ocal}{\mathcal O}

\newcommand{\Ncal}{\mathcal N}

\newcommand{\Ecal}{\mathcal E}

\newcommand{\ep}{\epsilon}

\newcommand{\beq}{\begin{equation}}
\newcommand{\eq}{\end{equation}}
\newcommand{\req}[1]{(\ref{#1})}

\catcode`\@=12 

\begin{document} 


\title{Non-Perturbative Quantum Geometry}

\pubnum{
SNUTP13-004
}
\date{November 2013}

\author{
Daniel Krefl$^a$   \\[0.2cm]
\it  $^a$ Center for Theoretical Physics, SNU, Seoul, South Korea\\
}

\Abstract{
The $\beta$-ensemble with cubic potential can be used to study a quantum particle in a double-well potential with symmetry breaking term. The quantum mechanical perturbative energy arises from the ensemble free energy in a novel large $N$ limit. A relation between the generating functions of the exact non-perturbative energy, similar in spirit to the one of Dunne-\"Unsal, is found. The exact quantization condition of Zinn-Justin and Jentschura is equivalent to the Nekrasov-Shatashvili quantization condition on the level of the ensemble. Refined topological string theory in the Nekrasov-Shatashvili limit arises as a large $N$ limit of quantum mechanics.}

\makepapertitle

\body

\version\versionno

\vskip 1em



\section{Introduction}

In this work our interest lies in the quantum mechanical problem of a spherically symmetric anharmonic oscillator at negative coupling. It has been known for some time that the energy levels and resonances of this quantum oscillator at negative coupling are equivalent to the energy levels and resonances of a different quantum problem at positive coupling. Namely, to a double-well potential with symmetry breaking term \cite{SZJ79,BV93}. The dynamics of the particle wave-function $\Psi(x)$ of the latter setup, equipped with proper boundary conditions, is captured by the time independent Schr\"odinger equation (\cf, \cite{ZJJ05a,ZJJ05b} and references therein)
\beq
\eqlabel{DWSBSE}
-\frac{\hbar^2}{2}\Psi''(x)+\left(\frac{1}{2}\left(x^2-\frac{1}{4}\right)^2-\hbar\, j\, x\right) \Psi(x)=\frac{\hbar\,\cale}{2}\Psi(x)\,,
\eq
with $j$ parameterizing the symmetry breaking term and $\cale$ denoting the energy. Special cases are $j=0$ and $j=-1$, which correspond under the additional rescaling $\cale\rightarrow 2\cale$ to the quantum system with ordinary double well, respectively, Fokker-Planck potential. 

In general, the energy $\cale$ can be split as 
$$
\cale=\cale_p(\Ncal,j;\hbar)+\cale_{np}(\Ncal,j;\hbar,\Lambda)\,,
$$
there $\cale_{p}$ denotes the perturbative part of the energy, \ie,
\beq
\eqlabel{Epert}
\Ecal_p(\Ncal,j;\hbar)=\sum_{n=0}^\infty E^{(n)}_p(\Ncal,j)\, \hbar^n\,,
\eq
with $E_p^{(n)}$ the expansion coefficients, while $\cale_{np}$ refers to contributions to the energy of non-perturbative origin. More precisely, the contribution of multi-instanton effects. The latter is taken to be of the form 
\beq
\eqlabel{1instDef}
\cale_{np}(\Ncal,j;\hbar)=\sum_{n=1}^\infty E_{np}^{(n)}(\Ncal,j;\hbar) \,\Lambda^n\,,
\eq
with $E_{np}^{(n)}(\Ncal,j;\hbar)$ denoting the $n$-instanton contribution as a series in $\hbar$ including possible singular terms, like for example contributions $\sim\log\hbar$ and/or $\sim 1/\hbar^c$ with $c$ some positive integer. We also defined $\Lambda:=e^{-\frac{1}{3\hbar}}$, which we refer to as instanton counting parameter. Note that both $E^{(n)}_p$ and $E^{(n)}_{np}$ are functions of the perturbative energy level $\Ncal$.

Astonishingly, it has been conjectured that the exact energy $\cale$, including all non-perturbative contributions, can be deduced from an `exact' quantization condition of the qualitative form \cite{JZZ82} (see also \cite{ZJJ03} for a brief summary)
\beq
\eqlabel{NPquantCond}
\Delta\left(B(\cale)\right)\sim \left(-\frac{2}{\hbar}\right)^{B(\cale)} e^{-A_p(\cale)}\, \Lambda\,,
\eq
with $A_p(\cale)$ and $B(\cale)$ energy dependent series in $\hbar$, which we refer to as generating functions of the non-perturbative energy. Formally, exact quantization conditions like \req{NPquantCond} can be derived from resurgence theory \cite{DDP97}. However, for this work it will not be necessary to consider resurgence, \ie, we will obtain \req{NPquantCond} more or less for free in our framework below. One should keep in mind that essentially the function $B$ is by definition the map between the non-perturbative energy level $\Ncal_{np}$ and the energy $\cale$ (\cf, \cite{JZZ82}), \ie,
\beq
\eqlabel{BquantCond}
\boxed{
B(\cale)= 2\Ncal_{np}+1+j}
\,.
\eq
The function $\Delta(B(\cale))$ occurring in \req{NPquantCond} refers to some combination of $B(\cale)$ dependent $\Gamma$-functions. The precise form of $\Delta$ (and the relative phase in \req{NPquantCond}) does not only depend on the form of the potential, but as well on the choice of boundary conditions. In contrast, the generating functions are solely determined by the potential. Usually, one uses the more natural combination $A:=\log\Lambda+A_p$ in \req{NPquantCond}. However, we prefer to work in terms of $A_p$, as it clearly illustrates that the right-hand side of \req{NPquantCond} is under expansion in $\Lambda$ of leading order $\Lambda^1$. Under expanding $B(\cale)=B(\cale_p)+\Ocal(\Lambda)$ (and noting that for $\Lambda\rightarrow 0$ one has that $\Ncal_{np}\rightarrow \Ncal$) one recovers from \req{BquantCond} the usual perturbative quantization condition $B(\cale_p)=2\Ncal+1+j$ (equivalently $\Delta(B(\cale_p))=0$). It is important to keep in mind that the statement above is that the map $B$ takes the same functional form for the full energy $\cale$ and the perturbative energy $\cale_p$.

There as it is not hard to calculate the function $B$ via WKB methods, the calculation of the function $A$ has been originally more involved. However, recently it has been discovered that the two generating functions $A$ and $B$ are not truly independent as a simple functional relation between them exists \cite{DU2013a}, at least for the quantum system \req{DWSBSE} under consideration, and as well for some other examples.

The purpose of this work is to report on an interesting observation linking quantum mechanics to $\beta$-ensembles and as well to refined topological string theory in the Nekrasov-Shatashvili limit. Recall that $\beta$-ensembles are defined as a generalization of usual matrix models via the partition function
\beq
\eqlabel{ZensembleDef}
Z(N,g_s):=\int[d\lambda] \Delta(\lambda)^{2\beta}\,e^{-\frac{\beta}{g_s} \sum_{i=1}^N W(\lambda_i)}\,,
\eq
with $\Delta(\lambda)$ the Vandermonde determinant $\Delta(\Lambda)=\prod_{i<j}^N(\lambda_i-\lambda_j)$ and $\beta$ some positive integer. We also assume that $W(\lambda)$ is a polynomial potential. At large $N$, with $S:= g_s\beta N$ fixed, the $\beta$-ensemble \req{ZensembleDef} calculates the partition function of the refined topological string on the corresponding Dijkgraaf-Vafa geometry \cite{DV09}. In particular, the Nekrasov-Shatashvili limit is defined as (a small $g_s$ expansion of $Z$ is implicit)
\beq
\eqlabel{FNSdef}
\Fcal_{NS}(S, g_s):=\lim_{\beta\rightarrow 0}\beta \log Z\left(S/(g_s\beta),\beta g_s\right)\,.
\eq
However, note that in the limit $\beta\rightarrow 0$ at large $N$ one may as well keep just $\Ncal:=\beta N$ fixed, leading to the partition function 
\beq
\eqlabel{FQdef}
\boxed{
\Fcal_Q(\Ncal, \hbar):=\lim_{\beta\rightarrow 0}\beta \log Z\left(\Ncal/\beta,\beta\hbar\right)
}\,.
\eq
Both $\Fcal_{NS}$ and $\Fcal_Q$ are at large $N$. However, one may also see $\Fcal_{NS}$ as a large $\Ncal$ limit of $\Fcal_Q$. The reason why the limit leading to $\Fcal_Q$ is of interest for us is that it makes contact with ordinary quantum mechanics, both on a perturbative and as well on a non-perturbative level. Therefore, we will refer to the limit \req{FQdef} as quantum mechanics limit of the $\beta$-ensemble. In particular, we will recover from the $\beta$-ensemble with a cubic potential $W(\lambda)$ the perturbative energy of the quantum mechanical problem \req{DWSBSE} (and so the map $B$), the generating function $A$ and as well the exact quantization condition \req{NPquantCond} ! 

The underlying relations can be qualitatively sketched as follows
\beq
\eqlabel{Sketch}
\begin{CD}
\beta\,{\rm ensembles} @>\substack{\beta\rightarrow 0,\, N\rightarrow\infty\\ \Ncal:=\beta N {\rm fixed}} >> {\rm Quantum}\, {\rm Mechanics} \\
@V\substack{N\rightarrow \infty\\ S:=g_s\beta N\,{\rm fixed}}VV	@VV\substack{ \Ncal\rightarrow\infty\\S:=\hbar \Ncal\,{\rm fixed}}V\\
{\rm Refined}\, {\rm Topological}\, {\rm Strings} @>>\beta\rightarrow 0>{\rm Quantum}\, {\rm Geometry}
\end{CD}
\eq
In order to avoid confusion one should note that there are actually two quantum special geometries in the game, related by the limiting procedure sketched in the right hand side of \req{Sketch}. One in terms of the $\Ncal$ variable and one in terms of $S$. The latter has been originally discovered in the Nekrasov-Shatashvili limit of refined topological strings (and gauge theory) \cite{MM09,ACDKV11}, while the existence of the former at finite $\Ncal$ is implied by this work. 

The outline is as follows. In section \ref{CubicSec} we will utilize the saddle-point approximation to calculate the free energy of a $\beta$-ensemble with cubic potential, and will observe that from the resulting free energy one can recover the perturbative energy $\cale_p$ of the quantum mechanical problem \req{DWSBSE} and so $B(\cale_p)$. In addition, we will find that the generating function $A(\cale_p)$ simply corresponds to the perturbative part of a combination of B-periods of the underlying (quantum) special geometry. In section \ref{NPquantGeometry}, we will discuss the non-perturbative side of the story. In particular, we will show that the `exact' quantization condition \req{NPquantCond} corresponds on the level of the ensemble to the Nekrasov-Shatashvili quantization condition, stated in \cite{NS09}. We conclude with a brief outlook in section \ref{outlook}. Appendix \ref{appA} collects a technical, but for section \ref{NPquantGeometry} important result. Namely, the expansion of the Gaussian $\beta$-ensemble free energy in the quantum mechanics limit defined in \req{FQdef}.

\section{The cubic ensemble}


\label{CubicSec}
\subsection{Saddle-point approximation}
\label{saddlepointapp}
Consider the eigenvalue ensemble \req{ZensembleDef} with cubic potential
\beq
\eqlabel{cubicPotential}
W(x)=\frac{1}{3}x^3-\frac{\delta}{4}x\,.
\eq
This cubic potential possesses the two critical points 
$$
x_*^\pm=\pm \frac{\sqrt{\delta}}{2}\,.
$$
We want to explicitly calculate the free energy of the eigenvalue ensemble \req{ZensembleDef} with potential \req{cubicPotential}. For that, we will perform a saddle-point approximation of the ensemble, which has been already discussed extensively in the literature. Therefore, we only need to sketch the basics, following \cite{KMT02,MS10,KW12}. 

Since the cubic \req{cubicPotential} has the two critical points $x_*^\pm$, we have to distribute $N^-$ eigenvalues around $x_*^-$ and $N^+$ eigenvalues around $x_*^+$ (with $N=N^-+N^+$), and consider a small fluctuation $y_i^{\pm}$ of the eigenvalue around the critical point it is located on, \ie,
\beq
\eqlabel{SaddlePointSplit}
(\lambda_1,\lambda_2,\dots,\lambda_{N})\rightarrow (x_*^- + y^-_1, \dots,x_*^- + y^-_{N_-},x_*^++  y^+_{1},\dots,x_*^++y^+_{N_+} )\,.
\eq
Effectively, this means that we write the eigenvalue ensemble as two eigenvalue ensembles coupled via a potential. 

The potential \req{cubicPotential} decomposes under \req{SaddlePointSplit} into
\beq
\eqlabel{Wsplit}
W(\lambda)\rightarrow\frac{\delta^{3/2}}{12}(N^--N^+)-\frac{\delta^{1/2}}{2}(S^-_2-S^+_2)+\frac{1}{3}(S^-_3+S^+_3) \,,
\eq
where we introduced $S^{\pm}_j:=\sum_{i=1}^{N^\pm} (y^{\pm}_i)^j$. The Vandermonde can be rewritten under \req{SaddlePointSplit} as 
\beq
\eqlabel{Vsplit}
\Delta(\lambda)\rightarrow \left(-\sqrt{\delta}\right)^{N^-N^+} \Delta(y^+)\, \Delta(y^-)\, \exp\left(-\sum_{l=1}^\infty\sum_{r=0}^l\frac{(-1)^{r}}{l \delta^{l/2}}\binom{l}{r}S^+_r S^-_{l-r} \right)\,.
\eq
Since $\Delta(\lambda)$ does not carry a $g_s$ dependence, and the potential $W(\lambda)$ carries an overall factor of $g_s^{-1}$, we observe from \req{Wsplit} that under a rescaling 
\beq
\eqlabel{CubicRescale}
y^\pm_i\rightarrow \left(\pm\frac{g_s}{\beta \sqrt{\delta}}\right)^{1/2} y^\pm_i\,,
\eq
(translating to $S^\pm_k \rightarrow \left(\pm\frac{g_s}{\beta \sqrt{\delta}}\right)^{k/2} S^\pm_k$) and subsequent expansion for small $g_s$, the partition function of the cubic can be turned into a sum of normalized gaussian correlators. This leads to a split of the partition function into three parts (\cf, \cite{MS10}), \ie,
\beq
\eqlabel{ZEEsplit}
Z= Z_{const}\times Z_{np} \times Z_{pert}\,.
\eq
The constant part $Z_{const}$ is the contribution independent of $S_k^\pm$ and factors out. Besides the obvious constant parts of \req{Wsplit} and \req{Vsplit}, we have also to include some additional factors due to the rescaling \req{CubicRescale} (originating from the two Vandermonds in \req{Vsplit} and the measure). Collecting all parts, one infers  
\beq
\eqlabel{cubicZconst}
\begin{split}
Z_{const}=&\,\delta^{N^-N^+\beta}\,(-1)^{(\beta N_-(N_--1)+N_-)/2}\, \left(\frac{g_s}{\beta\sqrt{\delta}}\right)^{(\beta N_-(N_--1)+\beta N_+(N_+-1)+N_-+N_+ )/2}  \\
&\times\exp\left(-\frac{\beta}{g_s}\left(\frac{\delta^{3/2}}{12}(N^--N^+)\right)\right)\,.
\end{split}
\eq

The perturbative factor $Z_{pert}$ is given by a sum of products of normalized gaussian correlators
$$
C^\pm_{k_1,k_2,\dots,k_m}(\beta) := \frac{1}{Z_{np}^\pm}\int [d\lambda] \Delta(\lambda)^{2\beta} \prod_{j=1}^m S^{\pm}_{k_j} \,e^{-\frac{1}{2}\sum_{i=1}^{N^\pm} \lambda_i^2}\,.
$$
(The reason being that the expansion in $g_s$ of $Z_{pert}$ pulls down sums of monomial insertions of $S_{k_j}^\pm$.) As worked out in \cite{MS10} (see also \cite{KW12}), the evaluation of such correlators is straight-forward, since $C^\pm_{0,0,\dots ,0}=N^\pm\times N^\pm\times\dots\times N^\pm$ and $C^\pm_{k_1,k_2,\dots,k_m}$ with $k_i\neq 0$ determined recursively by
$$
C^\pm_{n+1,k_1,k_2,\dots,k_m}=n(1-\beta)C^\pm_{n-1,k_1,\dots,k_m}+\beta\sum_{k=0}^{n-1}C^\pm_{n-k-1,k,k_1,\dots,k_m}+\sum_{j=1}^m k_jC^\pm_{k_1,\dots,k_j+n-1,\dots,k_m}\,.
$$
Finally, the gaussian normalization of the correlators is the origin of the factor 
$$
Z_{np}:=Z_{np}^+\times Z_{np}^-\,,
$$ 
in \req{ZEEsplit}. For later reference we recall that $Z_{np}^\pm$ is given by Mehta's integral
\beq
\eqlabel{cubicZnp}
Z_{np}^\pm = \int [d \lambda] \Delta(\lambda)^{2\beta}\, e^{-\frac{1}{2}\sum_{i=1}^{N^\pm} \lambda_i^2}=(2\pi)^{N^\pm/2}\prod_{n=1}^{N^\pm} \frac{\Gamma(1+n\beta)}{\Gamma(1+\beta)}\,.
\eq
Calculating $Z_{pert}$ order by order in $g_s$ and combining with the contributions \req{cubicZconst} and \req{cubicZnp} yields the partition function of the $\beta$-ensemble with potential \req{cubicPotential}, in the two cut phase.

\subsection{Perturbative quantum geometry}
We make the following claim. The quantum limit of the free energy, as defined in \req{FQdef}, of the $\beta$-ensemble with cubic potential \req{cubicPotential}, relates to the perturbative quantum mechanical energy $\cale_p$ occurring in \req{DWSBSE} via
\beq
\eqlabel{EpertFQ}
\boxed{
\cale_p=-4 \hbar \left.\frac{\partial \Fcal_Q(\Ncal,-\hbar/2)}{\partial\delta}\right|_{\delta=1}
}\,.
\eq
As the cubic is a two parameter model, we have to be more precise about what we mean with $\Ncal$, \ie, how we identify the energy level parameter $\Ncal$ in quantum mechanics with the number of eigenvalue parameters $N^\pm$ in the ensemble. We propose the identification of parameters
\beq
\eqlabel{NpmPara}
\boxed{
N^+=\aleph +\frac{s^+}{\beta}\,,\,\,\,\,\,N^-=-\aleph+\frac{s^-}{\beta}
}\,,
\eq
with $s^\pm$ constants to be detailed later. Correspondingly, we keep fixed at large $N$ in \req{FQdef}
$$
\Ncal^+:=\beta N^+\,,\,\,\,\,\,\Ncal^-:=\beta N^-\,,\,\,\,\,\, \Ncal:=\beta \aleph\,.
$$
An immediate implication of the identification of parameters above is that we have
\beq
\eqlabel{AperiodCombination}
\boxed{
\Ncal^+-\Ncal^-=2\Ncal+s^+-s^-:=\Pi_Q^A
}\,.
\eq
Hence, the so-defined combination of (quantum) A-periods  of the large $N$ geometry must actually be equal to the generating function $B(\cale_p)$, for proper choice of $s^\pm$, \ie, for $s^+-s^-=1+j$.

Note that one may see the relation in \req{EpertFQ} as an eigenvalue ensemble analog to Matone's relation in $\Ncal=2$ supersymmetric gauge theory \cite{M95}.\footnote{That a relation of this kind should hold for the cubic in the Nekrasov-Shatashvili limit was pointed out to the author by C. Vafa a couple of years ago.}  Qualitatively, the relation \req{EpertFQ} can be derived from a so-called brane insertion into the ensemble \req{ZensembleDef}, which naturally leads to a Schr\"odinger equation of the kind \req{DWSBSE}, following the Nekrasov-Shatashvili case discussed in \cite{ACDKV11}. This is as expected since $\left(W'(x)\right)^2=\left(x^2-\frac{\delta}{4}\right)^2$ corresponds to a double-well potential and the additional symmetry breaking term of order $\hbar$ can be traced back to originating from a shift of the kind \req{NpmPara}. However, we refrain to give the formal derivation here, as there appear to be some subtile normalization issues which we do not fully understand at the time being.

Let us instead explicitly verify the claim \req{EpertFQ}, making use of the ensemble free energy calculated as described in section \ref{saddlepointapp}. For that, note first that clearly $Z_{np}$ does not depend on $\delta$ (\cf, \req{cubicZnp}), and hence only $Z_{const}$ and $Z_{pert}$ do contribute to $\cale_p$. We obtain under the choice $s^+=1+j, s^-=0$ or $s^+=1, s^-=-j$ (both satisfy $s^+-s^-=1+j$), that \req{EpertFQ} yields the following leading terms of the perturbative quantum energy $\cale_p$ in a series in $\hbar$
\beq
\eqlabel{EpCubicResult}
\begin{split}
\Ecal_p(\Ncal)=&\,(1+j+2\Ncal)\sqrt{\delta}+(2+j^2+6\Ncal(1+\Ncal)+j(3+6\Ncal))\frac{\hbar}{\delta}\\
&-\frac{1}{2}(1+j+2\Ncal)(18+4j^2+34\Ncal(1+\Ncal)+17j(1+2\Ncal))\frac{\hbar^2}{\delta^{5/2}}+\Ocal(\hbar^3)\,,
\end{split}
\eq
Higher orders in $\hbar$ can be easily obtained from the perturbative $\beta$-ensemble calculation sketched in section \ref{saddlepointapp}, but are too lengthy to be explicitly shown here. Also note that in the ensemble the computational expense is not in powers of $\hbar$, but in powers of $\Ncal$. 

The perturbative energy \req{EpCubicResult} at $\delta=1$ obtained via the eigenvalue ensemble calculation precisely matches the result obtainable from the perturbative quantization condition. This can be easily checked via inverting the explicit expression for $B(\cale_p)$ given in \cite{ZJJ05b}. In particular, at $j=-1,\delta=1$ we can confirm the perturbative energy of the Fokker-Planck potential given in \cite{JZZ04} and at $j=0,\delta=1,\Ncal=0$ the ground state energy of the double well potential (both under $\hbar\rightarrow -\hbar$ and $\cale\rightarrow \cale/2$) \cite{ZJJ05a}.

It is instructive to further take the derivative $-\frac{1}{4\hbar}\frac{\partial \cale_p(\Ncal)}{\partial\Ncal}$ and perform an integration over $\delta$. This yields
\beq
\eqlabel{AnResult}
\begin{split}
A(\Ncal)&=-\frac{\delta^{3/2}}{3 \hbar}-\frac{3}{2}(1+j+2\Ncal)\log\delta\\
&- \frac{1}{6\delta^{3/2}}\left((35+21j^2-102\Ncal(1+\Ncal)-51j(1+2\Ncal)\right)\hbar \\
&+\frac{1}{4\delta^3}(1+j+2\Ncal)\left(139+41j^2+250\Ncal(1+\Ncal)+125j(1+2\Ncal)\right)\hbar^2+\Ocal\left(\hbar^3\right)\,.
\end{split}
\eq
But at $\delta=1$, and under usage of the perturbative quantization condition, this is nothing else than the other generating function, denoted as $A(\cale_p)$, occurring in the exact quantization condition of Zinn-Justin and Jentschura, as can be inferred by comparing to the explicit expansion for $A(\cale_p)$ given in \cite{ZJJ05b}. Hence, we deduce the relation
\beq
\eqlabel{EArelation}
\boxed{
\frac{1}{\hbar}\frac{\partial\cale_p}{\partial\Ncal}=-4 \frac{\partial A}{\partial\delta}=-4 \frac{\partial A_p}{\partial\delta}+2\frac{\delta^{1/2}}{\hbar}
}\,.
\eq
This is the new relation between the generating functions of the exact quantization condition promised in the abstract. However, one has to keep in mind that the relation \req{EArelation} holds for the $\delta$-dependent functions, hence, for the generating functions of a more general quantum problem than \req{DWSBSE} (more specifically, there the minima of the double well are parameterized by $\sqrt{\delta}$). The relation \req{EArelation} is similar in spirit to the relation found by Dunne-\"Unsal \cite{DU2013a}. In particular, the generating function $A(\cale_p,\delta)$ is completely determined by $B(\cale_p,\delta)$. 

In terms of the ensemble free energy $\Fcal_Q$ the relation \req{EArelation} reads
\beq
\eqlabel{PiBQDef}
\boxed{
\frac{\partial \Fcal_Q(\Ncal,\delta,-\hbar/2)}{\partial \Ncal}= A(\Ncal,\delta)+a(\Ncal):=\Pi^B_Q
}\,,
\eq
where $a(\Ncal)$ parameterizes the integration constant, \ie, a function of $\Ncal$ independent of $\delta$ and where we defined a (quantum) B-period $\Pi^B_Q$. Hence, one should see the relation \req{PiBQDef} as a (quantum) special geometry relation and \req{EArelation} being reminiscent thereof under the $\delta$ derivative.

\subsection{Large $\Ncal$ limit (of quantum mechanics)}
\label{LargeNpert}
So far we have shown that the quantum mechanical perturbative energy $\cale_p$ is essentially determined by a large $N$ limit of the ensemble free energy with $\beta N$ fixed (under a rescaling of $g_s$ and taking the derivative $\partial_\delta$). Correspondingly, the quantum mechanical energy is essentially the inverse of the quantum A-period $\Pi^A_Q$ of the ensemble large $N$ geometry.

From the ensemble point of view it is clear that there exists as well the usual t'Hooft limit with $S:=g_s \beta  N$ fixed (we work with a rescaled $g_s$ in the ensemble \req{ZensembleDef}, therefore the difference to the usual refined t'Hooft limit with $g_s \sqrt{\beta}N$ fixed), leading to refined topological string theory \cite{DV09} and at $\beta\rightarrow 0$, as in \req{FNSdef}, to the Nekrasov-Shatashvili limit thereof. 

What we learned above is that it is perfectly fine to take the limit $\beta\rightarrow 0$ with $\beta N$ fixed alone, leading to pure quantum mechanics. Of course, starting from quantum mechanics, we can then take as well $\hbar \rightarrow 0$ with $S=\hbar\,\Ncal$ fixed. This essentially yields by construction (up to integration in $\delta$ and rescaling of the coupling constant, \cf, \req{EpertFQ}, \req{FQdef} and \req{FNSdef}), the refined topological string free energy in the Nekrasov-Shatashvilli limit (since we ignore for the time being the integration constant $a(\Ncal)$, albeit without the non-perturbative contribution from $Z_{np}$ and without part of $Z_{const}$). For illustration, substituting $\Ncal\rightarrow S/\hbar$ into $\cale_p(\Ncal,-2\hbar)$ given in \req{EpCubicResult} yields
\beq
\eqlabel{EpS}
\begin{split}
-\frac{1}{4\hbar}\Ecal_p(S,-2\hbar)=&\,\frac{1}{4}\left(S\sqrt{\delta}-\frac{6S^2}{\delta}-\frac{68S^3}{\delta^{5/2}}-\frac{1500S^4}{\delta^4}+\Ocal\left(S^5\right)\right)\frac{1}{\hbar^2}\\
&+\frac{1}{8}\left(\sqrt{\delta}-\frac{12S}{\delta}-\frac{204S^2}{\delta^{5/2}} -\frac{6000S^3}{\delta^4}-\frac{213780S^4}{\delta^{11/2}}+\Ocal\left(S^5\right)\right)\frac{(1+j)}{\hbar}\\
& +\Ocal\left(\hbar^0\right)\,.
\end{split}
\eq
Up to integration in $\delta$ we recover at order $\hbar^{-2}$ the perturbative part of the tree-level free energy of the topological string on the cubic Dijkgraaf-Vafa geometry (on the anti-diagonal slice), see for instance \cite{CIV01}. Note that the order $\hbar^{-1}$ in \req{EpS} can be seen as originating from a shift $S\rightarrow S+\frac{(1+j)}{2}\hbar$ (\cf, \cite{KW10b, KW12})\,. Reversing the shift, we obtain an expansion of $\cale_p(S-(1+j)\hbar/2,-2\hbar)$ into even powers of $\hbar$ only, as is preferred for a refined topological string interpretation of the quantum mechanical energy. For instance, after reversing the shift, we have at order $\hbar^0$ (1-loop) 
$$
\frac{j^2-1}{8\delta}+\frac{(9j^2+19)S}{4\delta^{5/2}}+\frac{(129j^2-459)S^2}{2\delta^4}+\frac{(4455j^2-23405)S^3}{2\delta^{11/2}}+\Ocal\left(S^{4}\right)\,.
$$
At $j=0$ (corresponding to the ordinary double well) we recognize the 1-loop refined topological string free energy in the Nekrasov-Shatashvilli limit of the cubic on the anti-diagonal slice (\cf, \cite{ACDKV11}). Similarly, it can be checked that $j\neq 0$ corresponds to the refined topological string free energy under shifting $S_i\rightarrow S_i+\frac{1}{2}(1\pm j)g_s$ before going onto the anti-diagonal slice $S_2=-S_1=S$. Hence, from a topological string point of view, the symmetry breaking term in \req{DWSBSE} can be understood as a simple quantum shift of moduli. 

Making use of \req{PiBQDef} we conclude that
\beq
\eqlabel{PIBNS}
\widetilde\Pi_{NS}^B:=\frac{\partial \widetilde\Fcal_{NS}}{\partial S}= \frac{1}{\hbar}\left(A(S/\hbar,-2\hbar)+a(S/\hbar,-2\hbar)\right)\,,
\eq
where we denoted the Nekrasov-Shatashvili limit of the shifted refined topological string free energy as $\widetilde\Fcal_{NS}$, and the corresponding shifted quantum B-period as $\widetilde \Pi^B_{NS}$. Comparing \req{PiBQDef} and \req{PIBNS}, we deduce 
$$
\boxed{
\widetilde\Pi^B_{NS}=\frac{1}{\hbar}\Pi^B_{Q}(S/\hbar,-2\hbar)}
\,.
$$ 
That is, the shifted refined topological string quantum B-period is related to the quantum mechanical period just by the large $\Ncal$ substitution $\Ncal\rightarrow S/\hbar$, rescaling of the coupling constant and an overall factor.

\section{Non-perturbative quantum geometry}
\label{NPquantGeometry}
\subsection{Exact quantization}
Let us investigate the meaning of the integration constant $a(\Ncal)$ occurring in \req{PiBQDef} and \req{PIBNS} in more detail. Clearly, the integration constant is determined by the $\delta$ independent parts of $Z_{const}$ given in \req{cubicZconst} and $Z_{np}$ given in \req{cubicZnp} via the definition \req{FQdef}. While the $\beta\rightarrow 0$ limit of $\beta\log Z_{const}(-g_s/2)$ can easily be taken and gives a contribution
$$
(2\Ncal+s^+-s^-)\log  \left(-g_s/2\right) +(\Ncal -s^- -1/2)\pi\ii\,,
$$
the same limit applied to $\beta\log Z_{np}$ requires a bit more work. The details are worked out in appendix \ref{appA}, with final result stated in \req{aZnpPart}. Combining both contributions yields for the integration constant
\beq
a(\Ncal)=\left(\Ncal-s^--1/2\right)\pi\ii+(2\Ncal+s^+-s^-)\log \left(-g_s/2\right) +\log\frac{\Gamma(1+\Ncal+s^+)}{\Gamma(1-\Ncal+s^-)} \,.
\eq
Hence, exponentiating \req{PiBQDef} gives
\beq
\eqlabel{expPiBQ}
e^{-\frac{\partial \Fcal_Q}{\partial \Ncal}} = -\ii \left(-\frac{2}{g_s}\right)^{2\Ncal+s^+-s^-} e^{-A_p(\Ncal,s^+,s^-)-\pi\ii (\Ncal-s^-)} \,\frac{\Gamma(1-\Ncal+s^-)}{\Gamma(1+\Ncal+s^+)}\,\Lambda\,.
\eq
Recall from section \ref{LargeNpert} that $\frac{\partial \Fcal_Q}{\partial \Ncal}$ turns at large $\Ncal$ into $\hbar\frac{\partial \Fcal_{NS}}{\partial S}$ (up to a shift, which we neglect for convenience, as it is not of high relevance for the purpose of this section). However, the latter satisfies the so-called Nekrasov-Shatashvili quantization condition \cite{NS09,ACDKV11} 
\beq
\eqlabel{BSquant}
\exp\left(-\hbar\frac{\partial \Fcal_{NS}}{\partial S}\right) = 1\,.
\eq
One should note that the quantization condition \req{BSquant} is essentially the usual exact Bohr-Sommerfeld quantization condition, albeit with different integration contour, since
$$
\frac{\partial \Fcal_{NS}}{\partial S}=\Pi^B_{NS}=\oint_\calb \omega\,,
$$
where $\calb$ refers to a B-cycle in the large $\Ncal$ geometry and $\omega:=dx\,\partial_x\log \Psi(x)$ (with $\Psi(x)$ a brane partition function, \cf, \cite{ACDKV11}) is a quantum 1-form in the sense of \cite{MMM10,ACDKV11}. Essentially, the physical meaning of the condition \req{BSquant} is to impose uniqueness of the wave-function (brane partition function) under looping around the B-cycle. However, the uniqueness of the wave-function should hold both at large and finite $\Ncal$. Therefore, we learn that the same relation should hold for \req{expPiBQ}, \ie, $e^{-\Pi^B_Q}=e^{-\partial_\Ncal \Fcal_Q}=1$. In detail, we infer for the cubic
\beq
\eqlabel{quantCond}
\boxed{
\frac{\Gamma\left(1+\Ncal_{np}+s^+\right)}{\Gamma\left(1-\Ncal_{np}+s^-\right)}=\ii \left(-\frac{2}{g_s}\right)^{2\Ncal_{np}+s^+-s^-} e^{-A_p(\Ncal_{np},s^+,s^-)-\pi\ii (\Ncal_{np}-s^-+1) }\,\Lambda
}\,.
\eq
A remark is in order. If we impose the exact quantization condition \req{BSquant} the flat coordinate changes, \ie, $\Ncal\rightarrow\Ncal_{np}$. Correspondingly, the number of ensemble eigenvalues in a cut are now given by a non-perturbative quantum A-period in the large $N$ geometry. Therefore we substituted $\Ncal_{np}$ in \req{quantCond}. Comparing with \req{NPquantCond}, we observe that \req{quantCond} actually takes the form of an `exact' quantization condition in the sense of Zinn-Justin and Jentschura. It is remarkable that we obtain the quantization condition, which originally has been inferred from a multi-instanton calculation (or from resurgence), for free (strictly speaking not for free, as it translates to the derivation of the condition \req{BSquant}, which is however relatively simple, see \cite{ACDKV11}). On the level of the ensemble the multi-instanton contributions in quantum mechanics are essentially induced by the non-perturbative factor $Z_{np}$ (originating from the gaussian normalization). The precise relation to eigenvalue tunneling in the ensemble (naively one would expect such a relation) is not immediately clear (but we expect that this can be clarified along the lines of the WKB derivation of the exact quantization condition \cite{ZJJ03,ZJJ05a}, performed on the level of the ensemble) . 

Let us make use of the map \req{AperiodCombination} and further impose the condition $s^+-s^-=j+1$ to eliminate $s^+$ in \req{quantCond}. This yields
$$
\frac{\Gamma\left(\frac{1}{2}(1+j+B(\cale))+1+s^-\right)}{\Gamma\left(\frac{1}{2}(1+j-B(\cale))+1+s^-\right)}=\ii \left(-\frac{2}{g_s}\right)^{B(\cale)} e^{-A_p(\cale)-\pi\ii (B(\cale)-j-1)/2-\pi\ii(1+s^-)}\,\Lambda\,.
$$
Up to the additional term of $1+s^-$ in the $\Gamma$-functions, this exact quantization condition matches the exact quantization condition conjectured in \cite{ZJJ05a,ZJJ05b} for the resonances of the symmetric anharmonic oscillator.

The relation \req{quantCond} can be used to easily determine the full energy $\cale=\cale_p+\cale_{np}$, as we will discuss in more detail below. 

\subsection{Instanton expansion}
The exact quantization condition \req{quantCond} turns under using Euler's reflection formula $\Gamma\left(1-z\right)\Gamma\left(z\right)=\frac{\pi}{\sin \pi z}$ and condition \req{AperiodCombination} into
\beq
\eqlabel{QuantCondCos}
\frac{\sin\left(\frac{\pi(B(\cale)-s^+-s^-)}{2} \right)}{\pi}=\ii \left(-\frac{2}{g_s}\right)^{2\Ncal+s^+-s^-} \frac{e^{-A_p+\ii\pi(\Ncal-s^-)}}{\Gamma\left(\Ncal-s^-\right)\Gamma\left(1+\Ncal+s^+\right)}\,\Lambda\,.
\eq
We perform now the following little trick. We formally take $B(\cale_p+\cale_{np})$ and expand in $\Lambda$, which yields up to 2-instantons
\beq
\eqlabel{BdwLexpansion}
\begin{split}
B(\Ecal)=&\,(2\Ncal+s^+-s^-)+ \frac{\partial B(\Ecal_p)}{\partial \Ecal_p}E_{np}^{(1)}(\Ncal)\,\Lambda\\
&+\left( E^{(2)}_{np}(\Ncal) \frac{\partial B(\cale_p)}{\partial \cale_p}+\frac{1}{2}\left(E^{(1)}_{np}(\Ncal)\right)^2 \frac{\partial^2 B(\cale_p)}{\partial^2 \cale_p}\right)\Lambda^2+\Ocal\left(\Lambda^3\right)\,.
\end{split}
\eq
Note that we can parameterize this expansion as
\beq
\eqlabel{BnpDef}
B(\cale)=B(\cale_p)+\sum_{n=1}^\infty B^{(n)}(\cale_p) \,\Lambda^n\,,
\eq
with $B^{(n)}(\cale_p)$ denoting the coefficient of order $\Lambda^n$. The point being that we should actually see $B(\cale)$ as given in \req{BdwLexpansion} as a semi-classical limit of an exact quantum period, including non-perturbative corrections.

\paragraph{1-instanton}
Plugging \req{BdwLexpansion} into the left-hand side of \req{DWquanteq2} and expanding in $\Lambda$ gives (for $\Ncal-s^-$ integer)
$$
\frac{\sin\left(\frac{\pi(B(\cale)-s^+-s^-)}{2} \right)}{\pi}=\frac{(-1)^{\Ncal-s^-}}{2}\frac{\partial B(\Ecal_p)}{\partial \Ecal_p}E_{np}^{(1)}(\Ncal)\,\Lambda+\Ocal(\Lambda^2)\,.
$$
Since the right-hand side of \req{DWquanteq2} carries an overall factor of $\Lambda$, we immediately conclude that
\beq
\eqlabel{Enp1}
\boxed{
E_{np}^{(1)}(\Ncal)= \frac{\ii}{2\,(\Ncal+s^+)!(\Ncal-s^--1)!}\left(-\frac{2}{\hbar}\right)^{2\Ncal+s^+-s^-}e^{-A_p(\cale_p)}\, \frac{\partial \cale_p(\Ncal)}{\partial\Ncal}}   \,.
\eq
(This expression is not entirely novel, as a similar formula can be found for instance in \cite{ZJ02}). Using the explicit expansions \req{EpCubicResult} and \req{AnResult} we infer (at $\delta=1$)
$$
E^{(1)}_{np}(\Ncal)\sim 2+\left(\frac{53}{3}+j(23+7j)+(46+34j)\Ncal+34\Ncal^2 \right)\hbar+\Ocal\left(\hbar^2\right)\,.
$$
Up to the overall factor in \req{Enp1}, we see that $\frac{1}{2}E^{(1)}_{np}(0,-\hbar)$ matches at $j=-1$ the ground-state 1-instanton non-perturbative energy of the Fokker-Planck potential calculated in \cite{JZZ04}. 

In order to be able to compare to the double-well potential for $j=0$ we have actually to send $A_p\rightarrow A_p/2$ in \req{Enp1}. This yields (at $\delta=1$ and $j=0$)
\beq
\eqlabel{DW1instData}
\begin{split}
E_{np}^{(1)}(\Ncal)\sim&\, 1-\frac{1}{12}(71+174\Ncal+102\Ncal^2)\hbar\\
&+\frac{1}{288}(-6299-14172\Ncal-2112\Ncal^2+17496\Ncal^3+10404\Ncal^4)\hbar^2+\Ocal\left(\hbar^3\right)\,.
\end{split}
\eq
The explicit expansion \req{DW1instData} specializes to the known results for $E^{(1)}_{np}(0)$ and $E^{(1)}_{np}(1)$, as can be inferred by comparing to the expressions given in \cite{ZJJ05b} (again up to an overall factor).

Let us now invoke the perturbative relation \req{EArelation} to rewrite $E^{(1)}_{np}$ as
\beq
\eqlabel{E1np1dDelta}
E_{np}^{(1)}(\Ncal)\, \Lambda \sim -4\hbar\, e^{-A(\cale_p)} \frac{\partial A(\cale_p)}{\partial \delta}=4\hbar\frac{\partial}{\partial\delta}\left(e^{-A(\cale_p)}\right)\,.
\eq
It appears natural to promote the analog of Matone's relation \req{EpertFQ} to hold for the full energy $\cale$ such that if we expand the ensemble free energy $\Fcal_Q$ in powers of $\Lambda$ as
\beq
\eqlabel{FQNPdef}
\Fcal_Q(\Ncal)=\Fcal_Q^{(0)}(\Ncal)+\sum_{n=1}^\infty \Fcal^{(n)}_Q(\Ncal)\,\Lambda^n\,,
\eq
we deduce from \req{E1np1dDelta} that one should have at 1-instanton order
\beq
\eqlabel{FQ1instFinal}
\boxed{
\Fcal^{(1)}_Q(\Ncal) \sim -\hbar\, e^{-A_p(\Ncal)}
}
\,.
\eq
It is interesting to compare the proposed non-perturbative correction \req{FQ1instFinal} to the ensemble free energy in the limit \req{FQdef} to the 1-instanton correction due to instanton tunneling in usual 1-cut matrix models (see for instance \cite{MSW07}). Essentially, the implication being that one effect of the $\beta$-deformation on the non-perturbative sector is a simple change of the instanton action. That is, it seems natural that the instanton action becomes under the $\beta$-deformation equal to the quantum B-period (instead of the usual B-period).

Clearly, via considering the higher powers in $\Lambda$ of the expansion \req{BdwLexpansion} we can as well find closed analytic expressions for the higher $E_{np}^{(n>1)}$ out of \req{QuantCondCos} (and so for $\Fcal^{(n>1)}_Q$), though they will be more complicated. To give a flavor of how this works, we calculate in the following the 2-instanton sector of the double-well problem.

\paragraph{2-instantons: Double-well}
For that, we recall that the exact quantization condition of the double-well reads after invoking Euler's reflection formula \cite{ZJJ05a}
\beq
\eqlabel{DWquanteq2}
\frac{\cos\pi B(\cale)}{\pi}=\pm\ii \left(-\frac{2}{\hbar}\right)^{B(\cale)} \frac{e^{-A_p(\cale)/2}}{\sqrt{2\pi}\,\Gamma\left(\frac{1}{2}+B(\cale)\right)}\, \Lambda\,.
\eq
In retrospective, this explains why we had to rescale before $A_p$ to obtain \req{DW1instData}. The order $\Lambda^2$ of the left-hand side of \req{DWquanteq2} can be easily inferred from \req{BdwLexpansion}. The right-hand side at this order is a little bit more involved. For that, note that we have
\beq
\begin{split}
e^{- A_p(\cale)/2}&=e^{- A_p(\cale_p)/2}\left(1-\frac{1}{2} \frac{\partial A_p(\cale_p)}{\partial \cale_p} E^{(1)}_{np}(\Ncal)\,\Lambda+\Ocal\left(\Lambda^2\right)\right)\,,\\
\frac{1}{\Gamma\left(\frac{1}{2}+B(\cale)\right)}&=\frac{1}{\Ncal!}\left(1-\frac{\partial B(\Ecal_p)}{\partial \Ecal_p}E_{np}^{(1)}(\Ncal) \psi(1+\Ncal) \,\Lambda+\Ocal\left(\Lambda^2\right)\right)\,,
\end{split}
\eq
with $\psi(z)$ the digamma function.
Furthermore,
$$
\left(-\frac{2}{\hbar}\right)^{B(\cale)}=\left(-\frac{2}{\hbar}\right)^{B(\cale_p)}\left(1+\log\left(-\frac{2}{\hbar}\right) \frac{\partial B(\Ecal_p)}{\partial \Ecal_p}E_{np}^{(1)}(\Ncal)\,\Lambda +\Ocal\left(\Lambda^2\right)\right)\,.
$$
Collecting terms of order $\Lambda$ yields for the right-hand side of \req{DWquanteq2} at order $\Lambda^2$
$$
\pm\frac{\ii e^{- A_p(\cale_p)/2}}{\sqrt{2\pi}\,\Ncal!}\left(-\frac{2}{\hbar}\right)^{B(\cale_p)}E_{np}^{(1)}(\Ncal) \left(\left(\log\left(-\frac{2}{\hbar}\right)-\psi(\Ncal+1)\right)\frac{\partial B(\Ecal_p)}{\partial \Ecal_p}-\frac{1}{2}\frac{\partial A_p(\cale_p)}{\partial \cale_p} \right)\,.
$$
Hence, combining everything we deduce that
\beq
\eqlabel{DWEnp2}
\boxed{
\begin{split}
E^{(2)}_{np}(\Ncal)= \left(E^{(1)}_{np}(\Ncal)\right)^2&\left(\left(\log\left(-\frac{2}{\hbar}\right)-\psi(\Ncal+1)\right)\frac{\partial B(\Ecal_p)}{\partial \Ecal_p}\right. \\
&\left.-\frac{1}{2}\left(\frac{\partial A_p(\cale_p)}{\partial \cale_p}+\frac{\partial \cale_p(\Ncal)}{\partial \Ncal}\frac{\partial^2 B(\cale_p)}{\partial^2 \cale_p}\right) \right)
\end{split}
}\,.
\eq
Note that $\psi(\Ncal+1)=H_{\Ncal}-\gamma$, with $H_n$ the $n$th harmonic number and $\gamma$ the Euler-Mascheroni constant. $E^{(2)}_{np}(\Ncal)$ as given in \req{DWEnp2} can be easily evaluated for general level $\Ncal$. The first few leading orders in $\hbar$ are
\beq
\begin{split}
E^{(2)}_{np}(\Ncal)=\frac{1}{2\pi(\Ncal!)^2}\left(\frac{2}{\hbar}\right)^{2\Ncal+1}&\left(\left(\log\left(-\frac{2}{\hbar}\right)+\gamma-H_\Ncal\right)\right.\\
&\left.\left(1+\left(-\frac{56}{6}-23\Ncal-17\Ncal^2\right)\hbar+\Ocal\left(\hbar^2\right)\right)\right.\\
&\left. -\left(\frac{23}{2}+17\Ncal \right)\hbar -\left(\frac{35}{2}+51\Ncal+51\Ncal^2\right)\hbar^2\right.\\
&\left.+\Ocal\left(\hbar^3\right)\right)\,.
\end{split}
\eq
Setting $\Ncal=0$ and $\Ncal=1$ reproduces the expansions of $E_{np}^{(2)}$ given in \cite{ZJJ05b}.

\section{Outlook}
\label{outlook}
Let us wrap up the essential results of this work and discuss their implications. Though not formally derived, we learned from sections \ref{CubicSec} and \ref{NPquantGeometry} that there exists a large $N$ limit of the cubic $\beta$-ensemble in which the (quantum) A- and B-periods of the large $N$ geometry correspond to the $B$ and $A$ generating functions of the exact quantum mechanical energy. Furthermore, we observed that the exact quantization condition in quantum mechanics is in essence equivalent to the Nekrasov-Shatashvili quantization condition on the level of the ensemble. The solution of the exact quantization condition in quantum mechanics naturally leads to the definition of `exact' (or non-perturbative) quantum periods, see \req{BnpDef} and \req{FQNPdef}, and hence to a non-perturbative quantum geometry at large $N$. The non-perturbative corrections can be calculated analytically, even for general energy level $\Ncal$.

The refined topological string (perturbative) quantum geometry, arising in the Nekrasov-Shatashvili limit, can be recovered in a large $\Ncal$ limit from quantum mechanics, as sketched in \req{Sketch}. It is natural to expect that the correspondence extends to the non-perturbative sector. That is, the Nekrasov-Shatashvili quantization condition \req{BSquant} on the level of refined topological strings in the Nekrasov-Shatashvili limit, if invoked properly, yields a non-perturbative completion thereof (clearly, the perturbative part $\exp\left(\hbar\, \partial_S \Fcal^{pert}_{NS}\right)$  (\cf, \req{ZEEsplit}) introduces under expansion for small $S$ and $g_s$ a factor of order $\Lambda^1$, as in quantum mechanics). We suspect that the non-perturbative corrections captured by the exact quantization may relate to pair creation in the topological string gravi-photon background. One should note that the work \cite{ACDKV11} (and related works) mainly considered the perturbative quantization, essentially neglecting the true meaning of \req{BSquant}. This provides an explanation of the observation of \cite{KM13}, and actually may yield a more natural non-perturbative completion than proposed in \cite{KM13} (based on \cite{HMMO13} and references therein). More natural in the sense that the non-perturbative completion induced by \req{BSquant} is expected to not be given by the unrefined (or fully refined) topological string beyond tree-level (up to the trivial relation due to the fact that the Nekrasov-Shatashvili limit of the free energy counts BPS state degeneracies with respect to the total spin $j_L+j_R$, which one can decompose as a sum of individual degeneracies). 

It is also important to note that in this work we did not clarify a possible relation to eigenvalue tunneling in the ensemble, \ie, that the non-perturbative corrections captured by the exact quantization may actually correspond to eigenvalue tunneling in the ensemble (\req{FQ1instFinal} hints into this direction). 

We leave the answers to these and other questions to followup works. 

\acknowledgments{
D.K. likes to thank the BCTP of the University Bonn for hospitality during part of this project. The work of D.K. was supported in part by a Simons fellowship, by the Berkeley Center for Theoretical Physics and the National Research Foundation of Korea Grant No. 2012R1A2A2A02046739.
}

\appendix

\section{Calculation of non-perturbative contribution}
\label{appA}
In this appendix we are going to calculate the contribution of the non-perturbative part $Z_{np}$ to the integration constant $a(\Ncal)$ of section \ref{CubicSec}. The calculation goes as follows.

With the help of the integral representation 
\beq
\eqlabel{logGint}		
\log\Gamma(z)=\int_0^\infty \frac{dx}{x}\frac{1}{e^x-1}\left((z-1)(1-e^{-x})+e^{-x(z-1)}-1\right)\,,
\eq
one obtains from \req{cubicZnp}
\beq
\eqlabel{cubicFnp}
\begin{split}
\log Z_{np}^\pm =&\frac{N^\pm}{2} \log 2\pi -\log \beta!\\
&+ \int_{0}^\infty \frac{dx}{x(e^x-1)} \left(\frac{\beta}{2} N^\pm(N^\pm+1)(1-e^{-x})+\frac{e^{-\beta N^\pm x}(e^{\beta N^\pm x}-1)}{e^{\beta x}-1}-N^\pm\right)   \,.
\end{split}
\eq
After substituting $N^\pm\rightarrow \Ncal^\pm/\beta$ we can take the limit
\beq
\begin{split}
\lim_{\beta\rightarrow 0} \beta\log Z^\pm_{np}(\Ncal^\pm/\beta)=&\,\frac{\Ncal^\pm}{2}\log 2\pi\\
&+\int_0^\infty \frac{dx}{x(e^x-1)}\left(\frac{(\Ncal^\pm)^2}{2}(1-e^{-x}) +\frac{(1-e^{-\Ncal^\pm x})}{x}-\Ncal^\pm \right)\,.
\end{split}
\eq
Taking the $\partial_{\Ncal^\pm}$ derivative yields
$$
\partial_{\Ncal^\pm} \lim_{\beta\rightarrow 0} \beta\log Z^\pm_{np}(\Ncal^\pm/\beta)=\frac{1}{2}\log 2\pi+\int_0^\infty \frac{dx}{x}\left(\Ncal^\pm e^{-x }+\frac{e^{-\Ncal^\pm x}-1}{e^x-1}\right)\,.
$$
We define
$$
\Pi^\pm_{Q,np}=\partial_{\Ncal^\pm} \lim_{\beta\rightarrow 0} \beta\log Z^\pm_{np}(\Ncal^\pm/\beta)\,.
$$
Then,
$$
\Pi^B_{Q,np}= \Pi^+_{Q,np}-\Pi^-_{Q,np}=\int_0^\infty\frac{dx}{x}\left((\Ncal^+-\Ncal^-)e^{-x}+\frac{e^{-\Ncal^+ x}-e^{-\Ncal^- x}}{e^x-1}\right)\,.
$$
It remains to perform the integration. For that, we introduce a cutoff $\epsilon$ and take later the limit $\ep\rightarrow 0$. The integration over the first term is easy under invoking the standard integral
\beq
\eqlabel{ints1}
\begin{split}
\int_\ep^\infty\frac{dx}{x} e^{-x} &= -\log\ep-\gamma+\Ocal(\ep)\,.\\
\end{split} 
\eq
The integration over the second term is more involved. Fortunately, integrals of this kind have been evaluated for instance in appendix A of \cite{KS13}, from which we infer
$$
\int_\ep^\infty \frac{dx}{x} \frac{e^{-\Ncal}}{e^x-1}=\frac{1}{\ep}+(\Ncal+1/2)(\log\ep+\gamma)+\log\Gamma(1+\Ncal)-\frac{1}{2}\log 2\pi+\Ocal(\ep)\,.
$$ 
Hence,
\beq
\Pi^B_{Q,np}=\log\Gamma(1+\Ncal^+)-\log\Gamma(1+\Ncal^-)\,.
\eq
Making use of the identification \req{NpmPara} we conclude
\beq
\eqlabel{aZnpPart}
\boxed{
\Pi^B_{Q,np}=\log\Gamma\left(1+\Ncal+s^+\right)-\log\Gamma\left(1-\Ncal+s^-\right)
}\,.
\eq
It is interesting to note that the origin of the relative sign between the $\log$ terms in \req{aZnpPart} lies in the identification \req{NpmPara}, enforcing the combination $\Pi^+-\Pi^-$. In case of the opposite combination, we would have had, besides equal signs, an additional $\log 2\pi$ term.

\end{document}